\documentclass[aps,prd,twocolumn, superscriptaddress,showpacs,preprintnumbers,amssymb]{revtex4-1}

\usepackage{color}
\usepackage{graphicx} % Include figure files

\usepackage{multirow}
\usepackage{array}
\usepackage{pict2e}
\usepackage{hyperref}
\graphicspath{{ps}}
\usepackage{adjustbox}
\usepackage{rotating}
\usepackage{lineno}

\usepackage[fleqn]{amsmath}

\begin{document}
\title{\quad\\[0.5cm] Search for the $B \to Y(4260) K, ~Y(4260) \to J/\psi \pi^+\pi^-$ decays}

%\noaffiliation
\affiliation{University of the Basque Country UPV/EHU, 48080 Bilbao}
\affiliation{Beihang University, Beijing 100191}
%%%\affiliation{University of Bonn, 53115 Bonn}
\affiliation{Brookhaven National Laboratory, Upton, New York 11973}
\affiliation{Budker Institute of Nuclear Physics SB RAS, Novosibirsk 630090}
\affiliation{Faculty of Mathematics and Physics, Charles University, 121 16 Prague}
%%%\affiliation{Chiba University, Chiba 263-8522}
\affiliation{Chonnam National University, Kwangju 660-701}
\affiliation{University of Cincinnati, Cincinnati, Ohio 45221}
\affiliation{Deutsches Elektronen--Synchrotron, 22607 Hamburg}
%%%\affiliation{Duke University, Durham, North Carolina 27708}
\affiliation{University of Florida, Gainesville, Florida 32611}
%%%\affiliation{Department of Physics, Fu Jen Catholic University, Taipei 24205}
\affiliation{Key Laboratory of Nuclear Physics and Ion-beam Application (MOE) and Institute of Modern Physics, Fudan University, Shanghai 200443}
\affiliation{Justus-Liebig-Universit\"at Gie\ss{}en, 35392 Gie\ss{}en}
\affiliation{Gifu University, Gifu 501-1193}
\affiliation{II. Physikalisches Institut, Georg-August-Universit\"at G\"ottingen, 37073 G\"ottingen}
\affiliation{SOKENDAI (The Graduate University for Advanced Studies), Hayama 240-0193}
\affiliation{Gyeongsang National University, Chinju 660-701}
\affiliation{Hanyang University, Seoul 133-791}
\affiliation{University of Hawaii, Honolulu, Hawaii 96822}
\affiliation{High Energy Accelerator Research Organization (KEK), Tsukuba 305-0801}
\affiliation{J-PARC Branch, KEK Theory Center, High Energy Accelerator Research Organization (KEK), Tsukuba 305-0801}
\affiliation{Forschungszentrum J\"{u}lich, 52425 J\"{u}lich}
%%%\affiliation{Hiroshima Institute of Technology, Hiroshima 731-5193}
\affiliation{IKERBASQUE, Basque Foundation for Science, 48013 Bilbao}
%%%\affiliation{University of Illinois at Urbana-Champaign, Urbana, Illinois 61801}
\affiliation{Indian Institute of Science Education and Research Mohali, SAS Nagar, 140306}
\affiliation{Indian Institute of Technology Bhubaneswar, Satya Nagar 751007}
\affiliation{Indian Institute of Technology Guwahati, Assam 781039}
\affiliation{Indian Institute of Technology Hyderabad, Telangana 502285}
\affiliation{Indian Institute of Technology Madras, Chennai 600036}
\affiliation{Indiana University, Bloomington, Indiana 47408}
\affiliation{Institute of High Energy Physics, Chinese Academy of Sciences, Beijing 100049}
\affiliation{Institute of High Energy Physics, Vienna 1050}
\affiliation{Institute for High Energy Physics, Protvino 142281}
%%%\affiliation{Institute of Mathematical Sciences, Chennai 600113}
\affiliation{INFN - Sezione di Napoli, 80126 Napoli}
\affiliation{INFN - Sezione di Torino, 10125 Torino}
\affiliation{Advanced Science Research Center, Japan Atomic Energy Agency, Naka 319-1195}
\affiliation{J. Stefan Institute, 1000 Ljubljana}
%%%\affiliation{Kanagawa University, Yokohama 221-8686}
\affiliation{Institut f\"ur Experimentelle Teilchenphysik, Karlsruher Institut f\"ur Technologie, 76131 Karlsruhe}
%%%\affiliation{Kavli Institute for the Physics and Mathematics of the Universe (WPI), University of Tokyo, Kashiwa 277-8583}
\affiliation{Kennesaw State University, Kennesaw, Georgia 30144}
%%%\affiliation{King Abdulaziz City for Science and Technology, Riyadh 11442}
\affiliation{Department of Physics, Faculty of Science, King Abdulaziz University, Jeddah 21589}
\affiliation{Kitasato University, Sagamihara 252-0373}
\affiliation{Korea Institute of Science and Technology Information, Daejeon 305-806}
\affiliation{Korea University, Seoul 136-713}
%%%\affiliation{Kyoto University, Kyoto 606-8502}
\affiliation{Kyungpook National University, Daegu 702-701}
\affiliation{LAL, Univ. Paris-Sud, CNRS/IN2P3, Universit\'{e} Paris-Saclay, Orsay}
\affiliation{\'Ecole Polytechnique F\'ed\'erale de Lausanne (EPFL), Lausanne 1015}
\affiliation{P.N. Lebedev Physical Institute of the Russian Academy of Sciences, Moscow 119991}
\affiliation{Liaoning Normal University, Dalian 116029}
\affiliation{Faculty of Mathematics and Physics, University of Ljubljana, 1000 Ljubljana}
\affiliation{Ludwig Maximilians University, 80539 Munich}
\affiliation{Luther College, Decorah, Iowa 52101}
%%%\affiliation{Malaviya National Institute of Technology Jaipur, Jaipur 302017}
\affiliation{University of Malaya, 50603 Kuala Lumpur}
\affiliation{University of Maribor, 2000 Maribor}
\affiliation{Max-Planck-Institut f\"ur Physik, 80805 M\"unchen}
\affiliation{School of Physics, University of Melbourne, Victoria 3010}
\affiliation{University of Mississippi, University, Mississippi 38677}
\affiliation{University of Miyazaki, Miyazaki 889-2192}
\affiliation{Moscow Physical Engineering Institute, Moscow 115409}
\affiliation{Moscow Institute of Physics and Technology, Moscow Region 141700}
\affiliation{Graduate School of Science, Nagoya University, Nagoya 464-8602}
%%%\affiliation{Kobayashi-Maskawa Institute, Nagoya University, Nagoya 464-8602}
\affiliation{Universit\`{a} di Napoli Federico II, 80055 Napoli}
%%%\affiliation{Nara University of Education, Nara 630-8528}
\affiliation{Nara Women's University, Nara 630-8506}
\affiliation{National Central University, Chung-li 32054}
\affiliation{National United University, Miao Li 36003}
\affiliation{Department of Physics, National Taiwan University, Taipei 10617}
\affiliation{H. Niewodniczanski Institute of Nuclear Physics, Krakow 31-342}
\affiliation{Nippon Dental University, Niigata 951-8580}
\affiliation{Niigata University, Niigata 950-2181}
%%%\affiliation{University of Nova Gorica, 5000 Nova Gorica}
\affiliation{Novosibirsk State University, Novosibirsk 630090}
\affiliation{Osaka City University, Osaka 558-8585}
%%%\affiliation{Osaka University, Osaka 565-0871}
\affiliation{Pacific Northwest National Laboratory, Richland, Washington 99352}
\affiliation{Panjab University, Chandigarh 160014}
\affiliation{Peking University, Beijing 100871}
%%%\affiliation{University of Pittsburgh, Pittsburgh, Pennsylvania 15260}
\affiliation{Punjab Agricultural University, Ludhiana 141004}
%%%\affiliation{Research Center for Electron Photon Science, Tohoku University, Sendai 980-8578}
%%%\affiliation{Research Center for Nuclear Physics, Osaka University, Osaka 567-0047}
\affiliation{Theoretical Research Division, Nishina Center, RIKEN, Saitama 351-0198}
%%%\affiliation{RIKEN BNL Research Center, Upton, New York 11973}
%%%\affiliation{Saga University, Saga 840-8502}
\affiliation{University of Science and Technology of China, Hefei 230026}
\affiliation{Seoul National University, Seoul 151-742}
%%%\affiliation{Shinshu University, Nagano 390-8621}
\affiliation{Showa Pharmaceutical University, Tokyo 194-8543}
\affiliation{Soongsil University, Seoul 156-743}
%%%\affiliation{University of South Carolina, Columbia, South Carolina 29208}
%%%\affiliation{Stefan Meyer Institute for Subatomic Physics, Vienna 1090}
\affiliation{Sungkyunkwan University, Suwon 440-746}
\affiliation{School of Physics, University of Sydney, New South Wales 2006}
\affiliation{Department of Physics, Faculty of Science, University of Tabuk, Tabuk 71451}
\affiliation{Tata Institute of Fundamental Research, Mumbai 400005}
%%%\affiliation{Excellence Cluster Universe, Technische Universit\"at M\"unchen, 85748 Garching}
\affiliation{Department of Physics, Technische Universit\"at M\"unchen, 85748 Garching}
\affiliation{Toho University, Funabashi 274-8510}
%%%\affiliation{Tohoku Gakuin University, Tagajo 985-8537}
\affiliation{Department of Physics, Tohoku University, Sendai 980-8578}
\affiliation{Earthquake Research Institute, University of Tokyo, Tokyo 113-0032}
\affiliation{Department of Physics, University of Tokyo, Tokyo 113-0033}
\affiliation{Tokyo Institute of Technology, Tokyo 152-8550}
\affiliation{Tokyo Metropolitan University, Tokyo 192-0397}
%%%\affiliation{Tokyo University of Agriculture and Technology, Tokyo 184-8588}
%%%\affiliation{Utkal University, Bhubaneswar 751004}
\affiliation{Virginia Polytechnic Institute and State University, Blacksburg, Virginia 24061}
\affiliation{Wayne State University, Detroit, Michigan 48202}
\affiliation{Yamagata University, Yamagata 990-8560}
\affiliation{Yonsei University, Seoul 120-749}
% \author{A.~Abdesselam}\affiliation{Department of Physics, Faculty of Science, University of Tabuk, Tabuk 71451} % Tabuk
\author{R.~Garg}\affiliation{Panjab University, Chandigarh 160014} % Panjab
  \author{V.~Bhardwaj}\affiliation{Indian Institute of Science Education and Research Mohali, SAS Nagar, 140306} % IISERM
\author{J.~B.~Singh}\affiliation{Panjab University, Chandigarh 160014} % Panjab
\author{I.~Adachi}\affiliation{High Energy Accelerator Research Organization (KEK), Tsukuba 305-0801}\affiliation{SOKENDAI (The Graduate University for Advanced Studies), Hayama 240-0193} % KEK
% \author{K.~Adamczyk}\affiliation{H. Niewodniczanski Institute of Nuclear Physics, Krakow 31-342} % Krakow
  \author{J.~K.~Ahn}\affiliation{Korea University, Seoul 136-713} % Korea
  \author{H.~Aihara}\affiliation{Department of Physics, University of Tokyo, Tokyo 113-0033} % Tokyo
  \author{S.~Al~Said}\affiliation{Department of Physics, Faculty of Science, University of Tabuk, Tabuk 71451}\affiliation{Department of Physics, Faculty of Science, King Abdulaziz University, Jeddah 21589} % Tabuk
% \author{K.~Arinstein}\affiliation{Budker Institute of Nuclear Physics SB RAS, Novosibirsk 630090}\affiliation{Novosibirsk State University, Novosibirsk 630090} % BINP
% \author{Y.~Arita}\affiliation{Graduate School of Science, Nagoya University, Nagoya 464-8602} % Nagoya
  \author{D.~M.~Asner}\affiliation{Brookhaven National Laboratory, Upton, New York 11973} % BNL
% \author{H.~Atmacan}\affiliation{University of South Carolina, Columbia, South Carolina 29208} % SouthCarolina
  \author{V.~Aulchenko}\affiliation{Budker Institute of Nuclear Physics SB RAS, Novosibirsk 630090}\affiliation{Novosibirsk State University, Novosibirsk 630090} % BINP
  \author{T.~Aushev}\affiliation{Moscow Institute of Physics and Technology, Moscow Region 141700} % MIPT
% ...
% BINP
% \author{T.~Zivko}\affiliation{J. Stefan Institute, 1000 Ljubljana} % Ljubljana
% \author{A.~Zupanc}\affiliation{Faculty of Mathematics and Physics, University of Ljubljana, 1000 Ljubljana}\affiliation{J. Stefan Institute, 1000 Ljubljana} % Ljubljana
% \author{N.~Zwahlen}\affiliation{\'Ecole Polytechnique F\'ed\'erale de Lausanne (EPFL), Lausanne 1015} % Lausanne
\collaboration{The Belle Collaboration}

%%%%%%%%%%%%%%%%%%%%%%%%%%%%%%%%%%%%%%%%%%%%%%%%%%%%%%%
%Abstract
%%%%%%%%%%%%%%%%%%%%%%%%%%%%%%%%%%%%%%%%%%%%%%%%%%%%%5
\begin{abstract}
  We report the results of a search for the $B \to Y(4260) K, ~Y(4260)\to J/\psi\pi^+\pi^-$
  decays. This study is based on a data sample corresponding to an integrated luminosity of 711~fb$^{-1}$,
  collected at the $\Upsilon(4S)$ resonance with the Belle detector at
  the KEKB asymmetric-energy $e^+ e^-$ collider. We investigate the $J/\psi\pi^+\pi^-$ invariant mass distribution in the range
  4.0 to 4.6 GeV/$c^2$ using both $B^+ \to J/\psi \pi^+\pi^- K^+$ and $B^0 \to J/\psi \pi^+\pi^- K^0_S$ decays. We find excesses of events above the background levels, with significances of 2.1 and 0.9 standard deviations for charged and neutral $B \to Y(4260) K$ decays, respectively, taking into account the systematic uncertainties. These correspond to upper limits on the product of branching fractions, ${\cal B}(B^+ \to Y(4260) K^+) \times {\cal B}(Y(4260) \to J/\psi \pi^+ \pi^-) <1.4 \times 10^{-5}$ and ${\cal B}(B^0 \to Y(4260) K^0) \times {\cal B}(Y(4260) \to J/\psi \pi^+ \pi^-) < 1.7 \times 10^{-5}$ at the 90\% confidence level.
%=(9.4 \pm 3.0 ^{+0.7}_{-1.5}) \times 10^{-6}$, here the first (second) uncertainty is statstical (systematic).
  
\end{abstract}
\maketitle
%%%%%%%%%%%%%%%%%%%%%%%%%%%%%%%%%%%%%%%%%%%%%%%%%%%%%%%%%%%%%%%%%%%%%55
%Introduction
%%%%%%%%%%%%%%%%%%%%%%%%%%%%%%%%%%%%%%%%%%%%%%%%%%%%%%%%%%%%%%%%%%%%%%%5

{\renewcommand{\thefootnote}{\fnsymbol{footnote}}}
\setcounter{footnote}{0}

%\linenumbers
The $Y(4260)$ state, also known as $\psi(4260)$~\cite{pdg}, was first seen by the {B{\small{A}}B{\small{AR}}} collaboration in
2005~\cite{BaBar} in the initial-state radiation (ISR) process
$e^+ e^- \to \gamma_{\rm ISR} Y(4260), ~Y(4260)\to J/\psi \pi^+ \pi^-$
and confirmed by the Belle and CLEO collaborations
using the same process~\cite{belle,cleo}. The world average mass and decay width of the $Y(4260)$ are $(4230 \pm 8)$~MeV/$c^2$
and $(55 \pm 19)$~MeV~\cite{pdg}, respectively. Due to its observation in ISR production, the $J^{PC}$ of the $Y(4260)$ is expected to be $1^{--}$. The decay of $Y(4260)$ to $J/\psi\pi^+\pi^-$
indicates the presence of a $c\bar{c}$ pair among its quark constituents. However, its mass and
properties are not consistent with those expected for any of the $c\bar{c}$ states in the 
charmonium spectrum, which makes it problematic to assign the $Y(4260)$ to one of the
conventional $c\bar{c}$ states with $J^{PC} = 1^{--}$.

Attempts have been made to identify $Y(4260)$
as a candidate for a mixed state, which is an admixture of charmonium and tetraquark states~\cite{admix}, a hybrid charmonium state, which is a bound state of charmonium with a gluon~\cite{t_5}, a tetraquark~\cite{tetra}, a mesonic molecule~\cite{t_3, t_2, t_4}, or a charmonium baryonium~\cite{t_6}. The $Z_c(3900)^\pm$ state, which as it is charged makes it a natural tetraquark candidate, has been observed by the BESIII and Belle collaborations in the $J/\psi \pi^\pm$ invariant mass spectrum of the $e^+ e^- \to Y(4260) \to J/\psi \pi^+ \pi^-$ process~\cite{besiii,belle_Zc}, provides further evidence of the unconventional nature of the $Y(4260)$.

A mixed-state model, based upon a QCD sum-rule approach~\cite{theo}, suggests the possible interval on the product of the branching fractions of $B^+ \to Y(4260) K^+, ~Y(4260) \to J/\psi \pi^+ \pi^-$ to be in the range $3.0 \times 10^{-8} -  1.8 \times 10^{-6}$. The B{\small{A}}B{\small{AR}} collaboration has measured a signal for the charged $B$ decay with a statistical significance of 3.1 standard deviations ($\sigma$) based on a data sample of 211 $\rm fb^{-1}$ which contains $(232 \pm 3) \times 10^{6} B\bar{B}$ pairs~\cite{babar}. They set the upper limit at the 95\% confidence interval to be $\mathcal{B}(B^+ \to Y(4260)K^+) \times \mathcal{B}(Y(4260) \to J/\psi\pi^+\pi^-) < 2.9 \times 10^{-5}$. Further improvement is required on the precision of both the theoretical estimate and experimental measurement to elucidate the structure of $Y(4260)$. 

 Recently, two resonance structures have been observed by the BESIII collaboration in a fit to the cross section of the $e^+ e^- \to  J/\psi \pi^+ \pi^-$ process~\cite{besIII}. The resonance structures are interpreted as $Y(4260)$ and $Y(4360)$ with measured masses $(4222.0 \pm 3.1 \pm 1.4)$~MeV/$c^2$ and $(4320.0 \pm 10.4\pm 7.0)$~MeV/$c^2$, respectively. The measured $Y(4260)$ mass is not significantly lower than world average~\cite{pdg}, from which it deviates merely about 1$\sigma$, and the $Y(4360)$ has not yet been confirmed. We assume the presence of $Y(4260)$ only in the $J/\psi\pi^+\pi^-$ invariant mass region of interest as in the previous measurements~\cite{belle, BaBar, cleo} instead of adopting the search for the improved mass region.

In this paper, we report a search for $B \to Y(4260) K, ~Y(4260) \to J/\psi\pi^+\pi^-$~\cite{charge} decays based on a data sample corresponding to an integrated luminosity of 711 $\rm fb^{-1}$ which contains 
$(771.58 \pm 10.57) \times 10^6 ~B\bar{B}$ pairs, collected with the
Belle detector~\cite{detector}
at the KEKB asymmetric-energy $e^+e^-$ collider~\cite{kekb} operating at the 
$\Upsilon(4S)$ resonance. 

As the well established $B \to \psi(2S) K$ and $B \to X(3872) K$ decays have the same topology as the $B\to Y(4260)K$ decays, these decays are used as control samples to validate and calibrate Monte Carlo (MC) simulations. The signal simulation sample for each decay mode is generated 
using EvtGen~\cite{evtgen}. Here, the decays of $\psi(2S)$, $X(3872)$ and $J/\psi$ are specified to be $\psi(2S) \to J/\psi \pi^+\pi^-$, $X(3872) \to J/\psi \pi^+\pi^-$ and $J/\psi \to \ell^+\ell^-$, respectively, while $K_S^0$ decays generically~\cite{pdg}. All radiation effects are taken 
into account using PHOTOS~\cite{photos}. The detector response 
is simulated using GEANT3~\cite{geant}.

The charged tracks  used in the analysis are
 required to originate from the interaction point (IP) and have their point of closest 
approach to the IP within 3.5 cm along the beam 
axis and 1.0 cm in the plane transverse to the beam axis. 
Identification of charged pions and kaons are based on the information from the
aerogel Cherenkov counter system, time-of-flight scintillation counter (TOF) and central drift chamber. 
 All of the information is combined to form the pion (kaon) likelihood, $\mathcal{L}_{\pi}~(\mathcal{L}_{K})$, and the selections are made 
on the basis of the likelihood ratio 
$\mathcal {R}_{\pi(K)} = {\mathcal{L}_{\pi(K)}}/( {\mathcal{L}_{\pi} + \mathcal{L}_{K}})$. Charged pions (kaons) are identified requiring 
$\mathcal {R}_{\pi} ~(\mathcal {R}_{K})> 0.6$ with an
identification efficiency of 94\% (86\%)
and a misidentification rate of 7.5\% (4\%) for misidentifying kaon (pion) as a pion (kaon), respectively. These efficiencies and misidentification rates are determined using a control sample of $D^{*+} \to D^0(K^-\pi^+)\pi^+$ decays in the kinematic region of interest.

A $K_S^0\to \pi^+\pi^-$ candidate decay is reconstructed from a pair of oppositely charged tracks with a $\pi^+\pi^-$ invariant mass in the range 488 MeV/$c^2 < M_{\pi\pi}<$ 508 
MeV/$c^2$ ($\pm 4\sigma$ around the nominal $K_S^0$ mass~\cite{pdg}). 
The selected candidates are required to satisfy the criteria described in Ref.~\cite{fangks}.

Muon identification~\cite{muon} utilizes the track penetration depth and hit-distribution pattern in the $K_L^0$ and $\mu$ detector, which are combined to form the muon likelihood, $\mathcal{L}_\mu$, and the selection is made on the basis of the likelihood ratio $\mathcal{R}_\mu =  {\mathcal L_{\mu}}/({\mathcal{L}_{\mu}+\mathcal L_\pi+\mathcal L_K})$. Muons are identified requiring $\mathcal{R}_{\mu} > 0.1$ with an identification efficiency of 93\% and a misidentification rate of 3\% for misidentifying a pion as a muon.
Electron identification~\cite{electron} utilizes the electromagnetic shower shape and $E_{\rm ECL}/p$ ratio, where $E_{\rm ECL}$ is the energy deposition in electromagnetic calorimeter and $p$ is the track momentum, as well as the information used in the charged hadron identification, except that from the TOF. All the information is combined to form the electron likelihood ratio, $\mathcal{R}_{e}$. Electrons are identified requiring $\mathcal{R}_{e}> 0.01$.

A $J/\psi$ candidate is reconstructed in its decay mode 
$J/\psi \to \ell^{+}\ell^{-}$,
where $\ell$ stands for $e$ or $\mu$. In the $J/\psi \to e^+e^-$ mode, the energy loss due to bremsstrahlung photons is recovered by including the four-momenta of the photons detected within 0.05 radians around the electron or positron initial direction in the invariant 
mass calculation; this mode is, hereinafter, referred to as $J/\psi \to e^+ e^- (\gamma)$.
An invariant mass of a $J/\psi$ candidate
is required to be in the range 
$ 3.05~{\rm GeV}/c^2 \le M_{ee(\gamma)}  \le 3.13~{\rm GeV}/c^2$ or
$ 3.07~{\rm GeV}/c^2 \le M_{\mu\mu}  \le 3.13~{\rm GeV}/c^2$. %, respectively.
 The asymmetric interval is taken for $e^+e^- (\gamma)$ to include the radiative tail due to the imperfect energy loss recovery.
A vertex- and mass-constrained fit is performed to the
selected $J/\psi$ candidates in order to improve their momentum resolution.
 
The selected $J/\psi$ candidate is then combined with a $\pi^+\pi^-$ pair to form 
$\psi(2S)$, $X(3872)$, and $Y(4260)$ candidates, requiring the $J/\psi\pi^+\pi^-$ invariant mass, $M_{J/\psi\pi\pi}$, to be in the range $ 3.67~{\rm GeV}/c^2 \le M_{J/\psi\pi\pi}  \le 3.70~{\rm GeV}/c^2$, 
$ 3.835~{\rm GeV}/c^2 \le M_{J/\psi\pi\pi}  \le 3.910~{\rm GeV}/c^2$
and $ 4.0~{\rm GeV}/c^2 \le M_{J/\psi\pi\pi}  \le 4.6~{\rm GeV}/c^2$,
respectively.
To reconstruct a $B^+~(B^0)$ candidate,
a $K^+$ ($K_S^0$) candidate is combined with a $\psi(2S)$, $X(3872)$ or $Y(4260)$ candidate.

To identify the $B$ meson, two kinematic variables, 
the beam-constrained mass 
($M_{\rm bc} = \sqrt{(E_{\rm beam}/c^2)^2 - \sum_i (p_{i}^{*}/c)^2}$) and the energy difference ($\Delta E = \sum_i E_i^* - E_{\rm beam}$),
are used to discriminate the signal from the background. Here, $E_{\rm beam}$ is the beam energy and $p_{i}^*$ ($E_i^*$) is the momentum 
(energy) of the $i^{\rm th}$ final-state particle of the reconstructed signal candidate, where both are evaluated in the $e^+e^-$ center-of-mass (CM) frame.
The $B$ candidates with $M_{\rm bc} > $ 5.27 GeV/$c^2$ are
selected for further analysis.

Even after applying all the selection criteria, multiple $B$
candidates can be reconstructed from wrong combinations of the retained particles in an event. The mean number of $B$ candidates per event is found to be 1.6 (1.6), 1.7 (1.6) and 1.4 (1.2) for the charged (neutral) $B \to \psi(2S)(\to J/\psi\pi^+\pi^-) K$, $B \to X(3872)(\to J/\psi\pi^+\pi^-) K$ and $B \to Y(4260) K$ decays, respectively. In an event with multiple $B$ candidates,
we select the best candidate that has the smallest value of $\chi^2_{\rm BCS}$ = $\chi^2_{\rm vtx} + \chi^2_{M_{\rm bc}} + \chi^2_{J/\psi} (+\chi^2_{K_S^0})$,
where $\chi^2_{\rm vtx}$ represents the $\chi^2$ value obtained from a kinematic fit to the $B$ decay vertex for all the charged daughter particles, and the other $\chi^2$ values are evaluated using the reconstructed mass $M_i$ and its resolution $\sigma_i$ and the nominal mass $m_i^{\rm PDG}$~\cite{pdg} of the reconstructed meson $i$ as $\chi^2_i = [(M_i-m_i^{\rm PDG})/\sigma_i]^2$. Here, beam-constrained $M_{\rm bc}$ is used for the reconstructed mass in $\chi_{M_{\rm bc}}^2$, and $\chi_{K_S^0}^2$ is used only for the neutral $B$ decays. The reconstructed mass resolutions $\sigma_{M_{\rm bc}}$, $\sigma_{J/\psi}$, and $\sigma_{K_S^0}$ are evaluated in the $B\to\psi(2S) K$ decays to be 2.6 ${\rm MeV}/c^2$, 9.8 ${\rm MeV}/c^2$ and 1.6 ${\rm MeV}/c^2$, respectively. According to MC simulations, the best candidate selection identifies the true signal at rates of 76\% (72\%) for the charged (neutral) $B \to Y(4260) K$ decays. The same best candidate selection criterion are applied in the reconstruction of the control sample decays.

The dominant background comes from $e^+ e^- \to q\bar{q}$ ($q=u,d,s$ or $c$)
continuum events. To suppress this background, we utilize the difference in event topology between the isotropic distribution of particles in $B\bar{B}$ events and the jet-like collimation of particles in $q\bar{q}$ events by placing a requirement on the ratio of the second- and zeroth-order Fox-Wolfram moments~\cite{r2cut} to be less than 0.5.

Among the backgrounds from $B\bar{B}$ events, the main contribution
is expected to arise from inclusive $B$ decays to $J/\psi$.
To understand possible backgrounds, simulated sample of inclusive $B$ decays with a $J/\psi ~(\ell^+ \ell^-$) in the final state are studied; the sample corresponds to an integrated luminosity that is two orders of magnitude larger than that of data. No peaking structures are found in the $M_{J/\psi\pi\pi}$ signal regions of $B \to \psi(2S) K$, $B \to X(3872) K$, and
$B \to Y(4260) K$ decays. 
In order to check possible contributions from
non-$J/\psi$ sources, the $J/\psi$ mass sidebands
(2.54~GeV/$c^2<M_{J/\psi}<$ 2.72~GeV/$c^2$ and
3.32~GeV/$c^2< M_{J/\psi} <$ 3.50~GeV/$c^2$) are studied. The contributions are found to be negligible.

An unbinned extended maximum likelihood (UML) fit is performed to the 
$\Delta E$ distribution of each decay mode. The statistical weight for each candidate to be a signal decay is determined by using the $_{s}\mathcal{P}lot$ technique~\cite{splot}. The statistical weights
can be used to effectively subtract the combinatorial
background from the $M_{J/\psi\pi\pi}$
distribution of each decay mode. The signal yield of the intended resonance, then, can be extracted
from the weighted $M_{J/\psi\pi\pi}$ distribution, having a single background component of the non-resonant $B \to J/\psi\pi^+\pi^- K$ decays.

The $\Delta E$ variable is required to satisfy $\mathbin{-} 0.11 ~{\rm GeV} < \Delta E < 0.11 ~{\rm GeV}$ for the $B \to \psi(2S) K$, $X(3872)K$ and $Y(4260)K$ decay modes. The UML function used here is
\noindent
  \begingroup
  \small   
  \thinmuskip=\muexpr\thinmuskip*5/8\relax
  \medmuskip=\muexpr\medmuskip*5/8\relax  
\begin{equation}\label{eeq}
  \mathcal{L}(N_{\rm S},N_{\rm B}) = \frac{e^{-(N_{\rm S}+N_{\rm B})}}{N!}\prod_{i=1}^{N} [N_{\rm S}\times P_{\rm S}(x_i)+N_{\rm B}\times P_{\rm B}(x_i)]
\end{equation}
\endgroup
where $N$ is the total number of events, $N_{\rm S}$ ($N_{\rm B}$) is the number of signal (background) events, $P_{\rm S}$ ($P_{\rm B}$) is
the signal (background) probability density function (PDF) of the variable $x$, and the index $i$ runs over the total number of events. Here, the
signal refers to the charged or neutral $B \to J/\psi \pi^+ \pi^- K$ decays, the background refers to the combinatorial
background, and $x$ refers to the $\Delta E$ variable. The signal PDF is modeled by a sum of three Gaussians
for the $B \to Y(4260)K$ decay modes and by a sum of two Gaussians and a bifurcated Gaussian for the $B\to \psi(2S) K$ and $B \to X(3872) K$ modes. The mean and resolution of the core Gaussian are allowed to vary in the fit while the remaining shape and
normalization parameters are fixed to those obtained in the fit to the signal MC. The background
PDF is modeled by a first-order polynomial except for the $B \to X(3872)K$ decay mode, in which a
second-order polynomial is used. All parameters of the background PDF are allowed to vary in the fit.

\begin{figure}[htbp]
  \centering  
  \includegraphics[trim={0 0.8cm 0 1cm},width=0.45\textwidth]{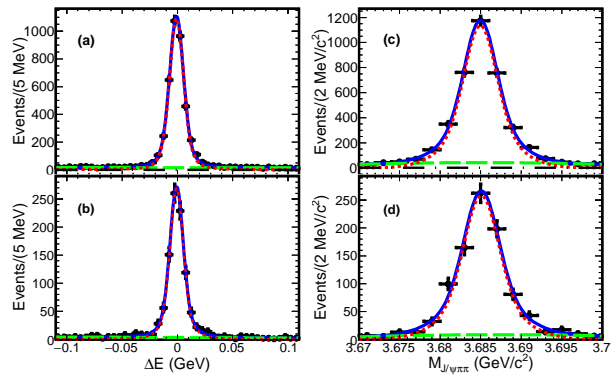}
  \caption{ Fit to the $\Delta E$ ((a) and (b)) and $_s\mathcal{P}lot$
    of $M_{J/\psi\pi\pi}$ ((c) and (d)) distributions for 
    $B^+ \to \psi(2S)(\to J/\psi \pi^+ \pi^-) K^+$ decays (top) 
    and $B^0 \to \psi(2S)(\to J/\psi\pi^+\pi^-) K_S^0$ decays (bottom),
    respectively. The curves show the fit functions for the signal (red dotted curve), background (green dashed curve) and their sum (blue solid line).}
  \label{fig:psi_data}
\end{figure}

\begin{figure}[ht!]
  \begin{center}
    \includegraphics[trim={0 1.5cm 0 1.5cm},width=0.45\textwidth]{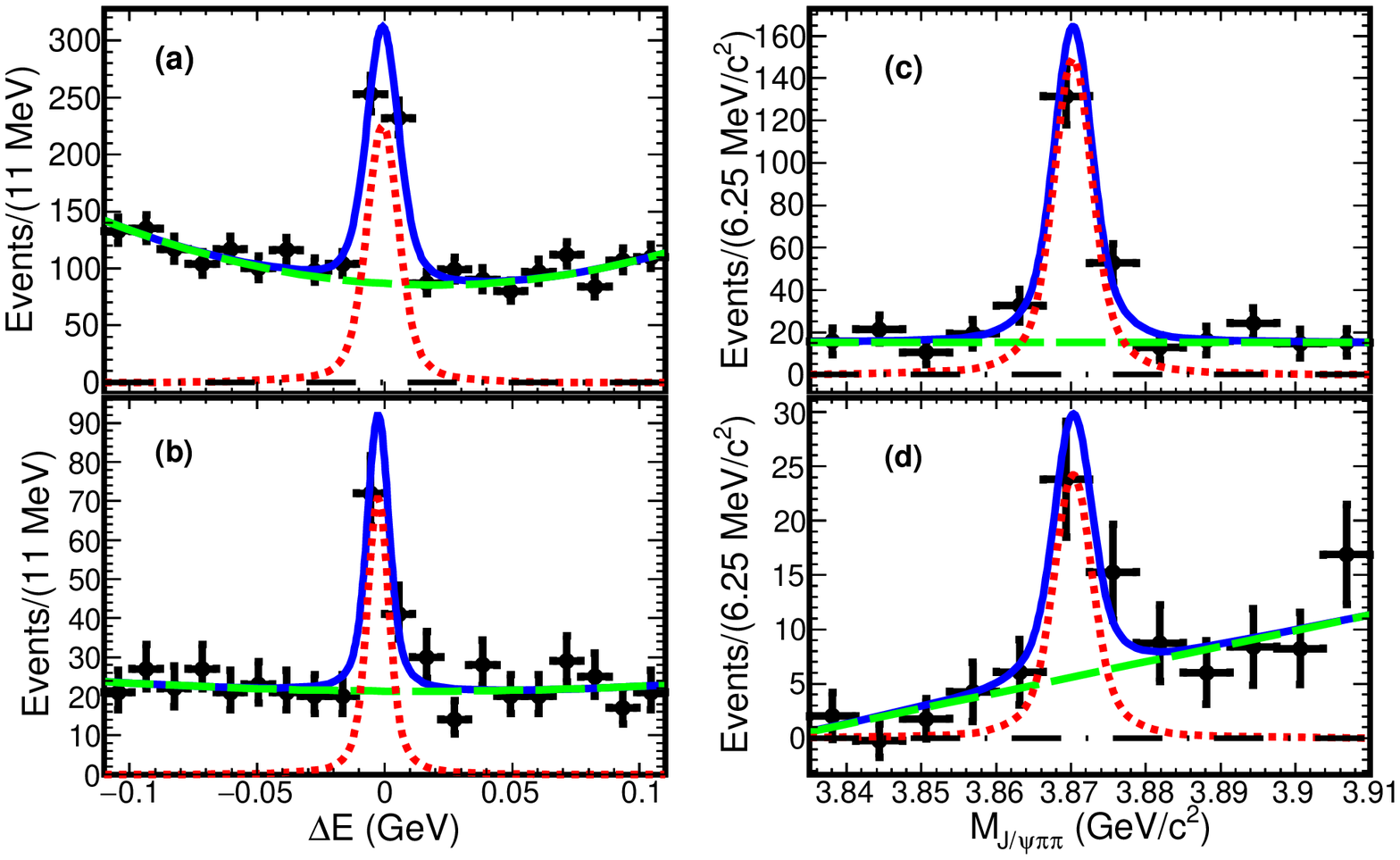}
  \end{center} 
  \caption{ Fit to the $\Delta E$ ((a) and (b)) and
    $_s\mathcal{P}lot$ of $M_{J/\psi\pi\pi}$ ((c) and (d)) distributions 
    for $B^+ \to X(3872)(\to J/\psi\pi^+\pi^-) K^+$ decays (top)
    and $B^0 \to X(3872)(\to J/\psi\pi^+\pi^-) K_S^0$ decays (bottom), respectively. Fit follow the same convention as Fig.~\ref{fig:psi_data}.}
    \label{fig:x_data}
\vspace{-0.5cm}
\end{figure}

\begin{figure}[htbp]
  \begin{center}
    \includegraphics[trim={0 1.5cm 0 1.1cm},width=0.45\textwidth]{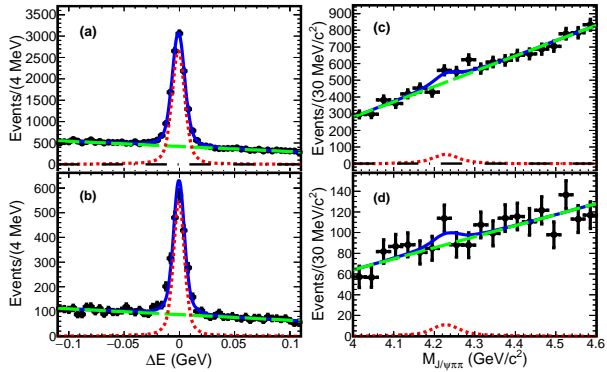}
  \end{center}
  \caption{  Fit to the $\Delta E$ ((a) and (b)) and $_s\mathcal{P}lot$ 
    of $M_{J/\psi\pi\pi}$ ((c) and (d)) distributions for 
    $B^+ \to Y(4260)(\to J/\psi\pi^+\pi^-) K^+$ decays (top)
    and $B^0 \to Y(4260)(\to J/\psi\pi^+\pi^-) K^0_S$ decays (bottom),
    respectively. Fit follow the same convention as Fig.~\ref{fig:psi_data}.}
  \label{fig:y_data}
\vspace{-0.5cm}
\end{figure}

The yields of the $B \to \psi(2S)K$, $X(3872)K$, and $Y(4260)K$ decays are extracted using independent UML
fits to the weighted $M_{J/\psi\pi\pi}$ distributions. Here, while the functional form of Eq.~\ref{eeq} is used to evaluate
the likelihood, the signal refers to the charged or neutral decay of $B \to \psi(2S)K$, $X(3872)K$, or $Y(4260)K$, the
background refers to the corresponding non-resonant $B \to J/\psi \pi^+ \pi^- K$ decay, and $x$ refers to the $M_{J/\psi\pi\pi}$
variable. The signal PDF is modeled by a sum of two Gaussians for the $B \to \psi(2S)K$ and $Y(4260)K$ decays
while an additional bifurcated Gaussian is used for the $B \to X(3872)K$ decays. The core Gaussian parameters for the $B \to \psi(2S)K$ and $B^+ \to X(3872)K^+$ decays are allowed to vary in the fit, while those for
the $B^0 \to X(3872)K^0$ and $B \to Y(4260)K$ decays are fixed to the values obtained in the fit to the signal
MC and calibrated with data; the calibration is based on the comparison of the shape parameters
between the data and simulation of the $B^+ \to X(3872)K^+$ decay. All the remaining shape and normalization parameters of the
signal PDF are fixed to those obtained in the fit to the signal MC. The background PDF is modeled by a first-order polynomial
except for the $B \to \psi(2S)K$ decay modes, in which a second-order polynomial is used. All parameters of
the background PDF are allowed to vary in the fit. The $\Delta E$ distributions, weighted $M_{J/\psi\pi\pi}$ distributions and projections of
their PDFs obtained from the fits are shown in Fig.~\ref{fig:psi_data},~\ref{fig:x_data}, and~\ref{fig:y_data} for the $B \to \psi(2S)K$, $X(3872)K$, and $Y(4260)K$ decay samples, respectively. The obtained signal yields of the $B \to \psi(2S)K$, and $B \to X(3872)K$,
$X(3872) \to J/\psi \pi^+ \pi^-$ decays are listed in Table~\ref{tab:br_data} and for $B \to Y(4260)K$, $Y(4260) \to J/\psi \pi^+ \pi^-$ decays are 
listed in Table~\ref{tab:br_y}.

\begin{table*}[htbp]
  \caption{\label{tab:br_data} Summary of the reconstruction efficiency ($\epsilon$), signal yield ($N_{\rm S}$) and branching fraction ($\mathcal{B}$) measured for the $B\to \psi(2S) K$ and $B\to X(3872)K, ~X(3872) \to J/\psi\pi^+\pi^-$ decays, together with the world average of the branching fraction ($\mathcal{B}_{\rm PDG}$)~\cite{pdg} for reference. Only the statistical uncertainty is included on the measured values of $N_{\rm S}$ and $\mathcal{B}$.} 
%  \vspace{-0.3 cm}
  \begin{center}
    \begin{tabular}{c| c |c|c |c}
      \hline
      \hline
      Decay &  $\epsilon$ (\%) & $N_{\rm S}$ & $\mathcal{B}$ & $\mathcal{B}_{\rm PDG}$\\
      \hline      
      $B^+\to \psi(2S) K^+$ & 16.8&$3481 \pm 95$&$(6.54 \pm 0.18) \times 10^{-4}$& $(6.21 \pm 0.23) \times 10^{-4}$  \\  
      $B^0 \to \psi(2S) K^0$ & 10.3 &$856 \pm 74$&$(5.25 \pm 0.45) \times 10^{-4}$&$(5.8 \pm 0.5) \times 10^{-4}$  \\ 
           $ B^+ \to X(3872) K^+, ~X(3872) \to J/\psi\pi^+\pi^-$& 22.2&$185 \pm 13$&$(9.07 \pm 0.64) \times 10^{-6}$&$(8.6 \pm 0.8) \times 10^{-6}$   \\
      
      $ B^0 \to X(3872) K^0, ~X(3872) \to J/\psi\pi^+\pi^-$ & 13.1&$29.9 \pm 6.2$&$(4.97 \pm 1.03) \times 10^{-6}$&$(4.3 \pm 1.3) \times 10^{-6}$  \\ \hline 
\hline
    \end{tabular}
  \end{center}
\vspace{-0.5cm}
\end{table*}

For the $B \to Y(4260)K$ decays, the statistical significance of the signal yield is evaluated using the likelihood
ratio as $\sqrt{-2 \rm ln(\mathcal{L}_0/\mathcal{L}_{\rm max})}$, where $\mathcal{L}_{\rm max}$ and $\mathcal{L}_0$ denote the maximum likelihood of the nominal fit and that of the fit with the null signal hypothesis. The statistical significances are evaluated to be 2.9$\sigma$ and
1.4$\sigma$ for the charged and neutral $B \to Y(4260)K$ decays, respectively.
The likelihood ratio is smeared with the systematic uncertainties, discussed later, and listed in
Table~\ref{tab:tot_sys}. The signal significances taking into account the systematic uncertainties are determined to be 2.1$\sigma$ and
0.9$\sigma$ for the charged and neutral $B \to Y(4260)K$ decays, respectively.

\par The branching fractions ($\mathcal{B}$) of the $B \to \psi(2S)K$ decays are obtained as 
$\mathcal{B}=N_{\rm S}/[N_{B\bar{B}} \times \epsilon \times f_K\times \mathcal{B}(\psi(2S) \to J/\psi\pi^+\pi^-)  \times \mathcal{B}(J/\psi \to \ell^+\ell^-)]$,
where $N_{\rm S}$ is the number of signal decays, $N_{B\bar{B}}$ is the number of $B\bar{B}$ events in the
data sample, and the branching fractions of the secondary decays are taken from Ref.~\cite{pdg}. Here, equal production of $B^+B^-$ and $B^0 \bar{B}^0$ pairs
from $\Upsilon(4S)$ decays is assumed. The reconstruction efficiency, $\epsilon$, is estimated from the signal MC
simulation, with the application of calibrations to account for discrepancies between the data and
signal MC related to particle identifications and $K^0_S$ reconstruction. These calibrations use dedicated control samples as
discussed later. The coefficient $f_K$ is introduced to translate the branching fractions for the final states
with $K^0_S$ into those for the ones with $K^0$ and set 1 and 0.5 for the charged and neutral $B \to \psi(2S)K$
decays, respectively. For the $B \to X(3872)K$ and $Y(4260)K$ decays, the branching fraction products are obtained in a similar manner as
\begin{align} \label{xyeq}
 \nonumber  \mathcal{B}(B^{+/0} \to R K^{+/0}) \times \mathcal{B}(R \to J/\psi \pi^+ \pi^-) = \\
  \frac{N_{\rm S}}{ N_{B\bar{B}} \times \epsilon \times f_K \times \mathcal{B}(J/\psi \to \ell^+ \ell^-)}
    \end{align}
where $R$ stands for the $X(3872)$ or $Y(4260)$ resonance. The obtained branching fractions of the $B \to \psi(2S)K$
decays and branching fraction products for the $B \to X(3872)K, ~X(3872) \to J/\psi\pi^+\pi^-$ decays are listed in
Table~\ref{tab:br_data} with the associated reconstruction efficiencies and signal yields. The obtained values agree well
with the world averages~\cite{pdg} and also with the previous Belle measurements~\cite{data}, indicating the
validity of the signal extraction procedure. The branching fraction products of the $B \to Y(4260)K, ~Y(4260) \to J/\psi\pi^+ \pi^-$ decays
as well as the associated reconstruction efficiencies and signal yields, are listed in Table~\ref{tab:br_y}.

%%%%%%%%%%%%%%%%%%%%%%%%%%%%%%%%%%%%%%%%%%%%%%%%%
\par With the absence of significant signals for the $B \to Y(4260)K$ decays, an upper limit (U.L.) is set on
each signal yield at the 90\% confidence level (C.L.) using a frequentist approach~\cite{freq}. The upper limits on the signal
yields at the 90\% C.L. ($N_{\rm S}^{UL}$) are found to be 259 and 84 events for the $B^+ \to Y(4260)K^+$ and $B^0 \to Y(4260)K^0_S$ decays, respectively. The upper limits on the branching fraction products are calculated using
Eq.~\ref{xyeq}, with $N_{\rm S}$ replaced by $N_{\rm S}^{UL}$ (systematic uncertainties are included in the upper limit calculation, as  will be described later in this paper).
The resulting upper limits are listed in
Table~\ref{tab:br_y}.

 \begin{table*}[hbt!]
  \caption{\label{tab:br_y} Summary of the reconstruction efficiency ($\epsilon$),
    signal yield ($N_{\rm S}$),
    signal significance ($\Sigma$) and the 90\% C.L. upper limit (U.L.) on the branching fraction for the
    $B^+ \to Y(4260) K^+$ and $B^0 \to Y(4260) K^0$ decays.
  }
  \begin{center}
    \begin{tabular}{c| c |c|c|c}
\hline
      \hline
      Decay & $\epsilon$ (\%)& $N_{\rm S}$ & $\Sigma$ ($\sigma$) & U.L.\\    \hline
      $ B^+ \to Y(4260) K^+, ~Y(4260) \to J/\psi \pi^+\pi^-$& 19.8&$179 \pm 53^{+55}_{-41}$ & 2.1 & $1.4 \times 10^{-5}$\\  
      $ B^0 \to Y(4260) K^0, ~Y(4260) \to J/\psi \pi^+\pi^-$&10.6& $39 \pm 28^{+~7}_{-31}$ & 0.9& $1.7 \times 10^{-5}$\\
       \hline
       \hline
    \end{tabular}
  \end{center}
  \vspace{-0.5cm}
 \end{table*}
 
 \par In order to improve the signal sensitivity, a simultaneous fit to the charged and neutral signal decays
 is performed keeping the fit procedure same as in the nominal fits for the individual signal decays, except for incorporating the constraint that $\mathcal{B}(B^+ \to Y(4260)K^+)/\mathcal{B}(B^0 \to Y(4260)K^0) = \mathcal{B}(B^+ \to \psi(2S) K^+) / \mathcal{B}(B^0 \to \psi(2S) K^0)$~\cite{sys}. The simultaneous fit for the $B \to Y(4260)K$ decays
 obtains $218 \pm 68$ signal events, where the quoted uncertainty is
statistical only. The combined
statistical significance of the $B \to Y(4260)K, ~Y(4260) \to J/\psi\pi^+ \pi^-$ decays is found to be 3.2$\sigma$, which reduces to 2.2$\sigma$ once systematic uncertainties are taken into account.
%After taking systematic uncertainties into account,
The simultaneous fit does not increase the significance of the $Y(4260)$ signal.% in comparison to the charged and neutral $B$ modes.

%%%%%%%%%%%%%%%%%%%%%%%%%%%%%%%%%%%%%%%%%%%%%%%%%%%%%%%%%%%%%%%%
\begin{table*}[!htbp]
   \centering
   \caption{\label{tab:tot_sys} Summary of the systematic uncertainties (\%) on the $B\to Y(4260)K$ branching fraction.}
     \begin{tabular}{|c|c|c|c |c| c |c|c|c|c|c|c|}
       \hline       
Source $\rightarrow$ & Tracking& \multicolumn{4}{c|}{Particle identification}&PDF & $Y(4260)$& Fit bias& $N_{B\bar{ B}}$&$\mathcal{B}(J/\psi \to \ell^+\ell^-)$& Total\\\cline{3-6}
 Decay $\downarrow$&& $K_S^0$& Kaon& Pion&Lepton &modeling&parameters&&&&\\
       \hline
  $B^+ \to Y(4260)K^+$ &    1.8&-& 0.9&1.3&1.2&$_{-11.1}^{+8.0}$& $^{+29.0}_{-19.5}$&4.3&1.4&0.4& $_{-23.0}^{+30.5}$\\
     $B^0 \to Y(4260) K_S^0$&  2.1&0.7&-&1.3&1.2& $^{+7.7}_{-77.0}$&$^{+14.0}_{-17.3}$&6.5&1.4&0.4& $_{-79.2}^{+17.5}$\\
      Simultaneous&  1.9&0.2&0.7&1.3&1.2&      $^{+5.3}_{-15.3}$&$^{+25.0}_{-18.0}$&4.8&1.4&0.4&$^{+26.2}_{-24.3}$\\
       \hline
     \end{tabular}
%\vspace{-0.2cm}
\end{table*}

All the systematic uncertainties are summarized in Table~\ref{tab:tot_sys}. The tracking efficiency in MC simulation is calibrated using a control sample of $D^* \to \pi D^0,~ D^0 \to  \pi^+\pi^- K_S^0, ~K_S^0 \to \pi^+\pi^-$ decays, and
the uncertainty on the calibration factor is 0.35\% per track. The calibration factor for
the $K^0_S$ reconstruction efficiency is obtained using $D^{*\pm} \to D^0(\to K_S^0\pi^0)\pi^\pm$ decays with an
uncertainty of 0.7\%. For the particle identification efficiencies, the calibration factors are obtained using
the dedicated control samples mentioned earlier, and the resulting systematic uncertainty is 0.9\% and 1.3\% for kaon and pion identification, respectively. The dominant systematic uncertainties are due to the PDF modeling,
and the values of the $Y(4260)$ mass and decay width~\cite{pdg} assumed in the fit. The changes on the signal yield from the nominal one
due to the uncertainty in the PDF modeling is estimated by varying each of the fixed parameters
independently by $\pm 1 \sigma$.
The corresponding changes due to the uncertainties on the $Y(4260)$
mass and decay width are estimated by separately applying the variation in the signal PDF based on the
alternative signal MC simulations, which are generated varying each of the mass and decay width in
the same manner. The resulting changes are added in quadrature.
The uncertainty in the PDF
modeling for the $B^0 \to Y(4260) K^0_S$ decay gives an exceptionally large systematic uncertainty of 77.0\%.
This is due to the systematic uncertainty associated with the background PDF modeling. The fit procedures are validated in fully simulated MC experiments with ensembles of signal and inclusive $B$ decays involving $J/\psi$.
The small biases of 4.3\%--4.8\% seen in the validation are taken as systematic uncertainties. The
uncertainties on $N_{B\bar{B}}$ and $\mathcal{B}(J/\psi \to \ell^+ \ell^-$), 1.4\% and 0.4\%, respectively, are also included in the systematic
uncertainties. The total systematic
uncertainties are estimated to be $_{-23.0}^{+30.5}$\%, $_{-79.2}^{+17.5}$\% and  $_{-24.3}^{+26.2}$\% on the results for
the charged, neutral and combined $B \to Y(4260)K, ~Y(4260) \to J/\psi\pi^+ \pi^-$ decays, respectively, by adding
all the sources in quadrature.

In summary, a search for the $B \to Y(4260)K$, $Y(4260) \to J/\psi\pi^+ \pi^-$ decays is performed using $B\bar{B}$ pairs collected at the $\Upsilon(4S)$ resonance by the Belle experiment at the KEKB. The observed signal yields are $179\pm 53 ^{+55}_{-41}$ events and $39 \pm 28 ^{+~7}_{-31}$ events for the charged and neutral $B \to Y(4260)K, ~Y(4260) \to J/\psi\pi^+ \pi^-$ decays, respectively, from fits to the individual decay samples; the first and second
uncertainties are statistical and systematic, respectively. The signal significances are obtained to be
$2.1\sigma$ and $0.9\sigma$ for the charged and neutral decays, respectively, taking into account the systematic uncertainties associated with the signal
extraction. %The branching fraction products are measured to be $\mathcal{B}(B^+
In the absence of
any significant signals, the upper limits on the branching fraction products at the 90\% C.L. are determined to be
$1.4 \times 10^{-5}$ and $1.7 \times 10^{-5}$ for the charged and neutral decays, respectively, taking into account the systematic uncertainties.
\par The obtained results give the most stringent upper limits, to date, on the branching fraction products
of the charged and neutral $B \to Y(4260)K, ~Y(4260) \to J/\psi\pi^+ \pi^-$ decays. The upper limits on the branching fraction products at the 95\% C.L. are also determined and are $1.56 \times 10^{-5}$ and $2.16 \times 10^{-5}$ for the charged and neutral decays, respectively. The upper limit for the charged
decay is consistent with the 95\% confidence interval set by the B{\small{A}}B{\small{AR}} collaboration~\cite{babar} and the one
for the neutral decay is given for the first time. While an excess of events above background is seen, improved measurements with a larger data sample are
demanded to establish signals and to elucidate the nature of the $Y(4260)$ state.

\par We thank the KEKB group for the excellent operation of the
accelerator; the KEK cryogenics group for the efficient
operation of the solenoid; and the KEK computer group, and the Pacific Northwest National
Laboratory (PNNL) Environmental Molecular Sciences Laboratory (EMSL)
computing group for strong computing support; and the National
Institute of Informatics, and Science Information NETwork 5 (SINET5) for
valuable network support.  We acknowledge support from
the Ministry of Education, Culture, Sports, Science, and
Technology (MEXT) of Japan, the Japan Society for the 
Promotion of Science (JSPS), and the Tau-Lepton Physics 
Research Center of Nagoya University; 
the Australian Research Council including grants
DP180102629, % Sevior
DP170102389, % Varvell
DP170102204, % Yabsley
DP150103061, % Urquijo
FT130100303; % Urquijo;
Austrian Science Fund under Grant No.~P 26794-N20;
the National Natural Science Foundation of China under Contracts
No.~11435013,  %Zhen-An Liu
No.~11475187,  %Chang-Zheng Yuan
No.~11521505,  %Chang-Zheng Yuan
No.~11575017,  %Cheng-Ping Shen
No.~11675166,  %Wen-Biao Yan
No.~11705209;  %Yi-Ming Li
Key Research Program of Frontier Sciences, Chinese Academy of Sciences (CAS), Grant No.~QYZDJ-SSW-SLH011; % Chang-Zheng Yuan
the  CAS Center for Excellence in Particle Physics (CCEPP); %Chang-Zheng Yuan, 
the Shanghai Pujiang Program under Grant No.~18PJ1401000;  %Tao Luo
the Ministry of Education, Youth and Sports of the Czech
Republic under Contract No.~LTT17020;
the Carl Zeiss Foundation, the Deutsche Forschungsgemeinschaft, the
Excellence Cluster Universe, and the VolkswagenStiftung;
the Department of Science and Technology of India; 
the Istituto Nazionale di Fisica Nucleare of Italy; 
National Research Foundation (NRF) of Korea Grants
No.~2015H1A2A1033649, No.~2016R1D1A1B01010135, No.~2016K1A3A7A09005
603, No.~2016R1D1A1B02012900, No.~2018R1A2B3003 643,
No.~2018R1A6A1A06024970, No.~2018R1D1 A1B07047294; Radiation Science Research Institute, Foreign Large-size Research Facility Application Supporting project, the Global Science Experimental Data Hub Center of the Korea Institute of Science and Technology Information and KREONET/GLORIAD;
the Polish Ministry of Science and Higher Education and 
the National Science Center;
the Grant of the Russian Federation Government, Agreement No.~14.W03.31.0026; % from 15.02.2018;
the Slovenian Research Agency;
Ikerbasque, Basque Foundation for Science, Spain;
the Swiss National Science Foundation; 
the Ministry of Education and the Ministry of Science and Technology of Taiwan;
and the United States Department of Energy and the National Science Foundation.

%%%%%%%%%%%%%%%%%%%%%%%%%%%%%%%%%%%%%%%%%%%%%%%%%%%%%%%%%%%%%%%%%%%%%
%                               Bibliography
%%%%%%%%%%%%%%%%%%%%%%%%%%%%%%%%%%%%%%%%%%%%%%%%%%%%%%%%%%%%%%%%%%%%%%

%\end{multicols}

\begin{thebibliography}{99}

\bibitem{pdg}
  M. Tanabashi {\it et al.} (Particle Data Group), Phys. Rev. D \textbf{98}, 030001 (2018).

\bibitem{BaBar}
B. Aubert {\it et al.} (B{\small{A}}B{\small{AR}} Collaboration), Phys. Rev. Lett. \textbf{95}, 142001 (2005).
\bibitem{belle}
C. Z. Yuan {\it et al.} (Belle Collaboration), Phys. Rev. Lett. \textbf{99}, 182004 (2007).
\bibitem{cleo}
Q. He {\it et al.} (CLEO Collaboration), Phys. Rev. D \textbf{74}, 091104(R) (2006).

%\bibitem{y4008}
%Z. Q. Liu {\it et al.} (Belle Collaboration), Phys. Rev. Lett. \textbf{110}, 252002 (2013).
%C. Patrignani {\it et al.} (Particle Data Group) Chin. Phys. C \textbf{40}, 100001 (2016) and 2017 update.

%\bibitem{psi3d}
 % S. Godfrey and N. Isgur, Phys. Rev. D \textbf{32}, 189 (1985).
\bibitem{admix}
J. M. Dias, R. M. Albuquerque, M. Nielsen and C. M. Zanetti, Phys. Rev. D \textbf{86}, 116012 (2012).
\bibitem{t_5}
S. L. Zhu {\it et al.} Phys. Lett. B \textbf{625}, 212 (2005).
\bibitem{tetra}
  L. Maiani, V. Riquer, F. Piccinini and A. D. Polosa Phys. Rev. D \textbf{72}, 031502 (2005).

%\bibitem{t_2}
%G. J. Ding {\it et al.} Phys. Rev. D \textbf{79}, 014001 (2009).

\bibitem{t_3}
X. Liu, X. Q. Zeng and X. Q. Li, Phys. Rev. D \textbf{72}, 054023 (2005).

\bibitem{t_2}
G. J. Ding,  Phys. Rev. D \textbf{79}, 014001 (2009).


\bibitem{t_4}
A. Martinez Torres, K. P. Khemchandani, D. Gamermann, and E. Oset,
Phys. Rev. D \textbf{80}, 094012 (2009).

%\bibitem{t_4}
%A. M. Torres, K. P. Khemchandani, D. Gamermann, E. Oset {\it et al.} Phys. Rev. D \textbf{80}, 094012 (2009).

\bibitem{t_6}
C. F. Qiao {\it et al.} Phys. Lett. B \textbf{639}, 263 (2006).
\bibitem{besiii}
M. Ablikim {\it et al.} (BESIII Collaboration), Phys. Rev. Lett. \textbf{110}, 252001 (2013).


\bibitem{belle_Zc}
Z. Q. Liu {\it et al.} (Belle Collaboration), Phys. Rev. Lett. \textbf{110}, 252002 (2013).

\bibitem{theo}
R. M. Albuquerque, M. Nielsen, C. M. Zanetti {\it et al.} Phys. Lett. B \textbf{747}, 83 (2015).
%\bibitem{theo1}
 % J. M. Dias, R. M. Albuquerque, M. Nielsen and C. M. Zanetti, Phys. Rev. D \textbf{86}, 116012 (2012)
\bibitem{babar}
B. Aubert {\it et al.} (B{\small{A}}B{\small{AR}} Collaboration), Phys. Rev. D \textbf{73}, 011101(R) (2006).

\bibitem{besIII}
M. Ablikim {\it et al.} (BESIII Collaboration), Phys. Rev. Lett. \textbf{118}, 092001 (2017).

%\bibitem{line}
%  M. Cleven, Q. Zhao, Phys. Lett. B \textbf{768}, 52 (2017).
\bibitem{charge}
  Charge-conjugate modes are implied throughout the paper.
\bibitem{detector}
A. Abashian {\it et al.} (Belle Collaboration), 
Nucl. Instrum. Methods Phys. Res., Sect. A {\bf 479}, 117 (2002);
also see detector section in J. Brodzicka {\it et al.}, 
Prog. Theor. Exp. Phys. {\bf 2012}, 04D001 (2012).

\bibitem{kekb}
S. Kurokawa and E. Kikutani, Nucl. Instrum. Methods Phys. Res., Sect. A {\bf 499}, 1 (2003), 
and other papers included in this volume;
T. Abe {\it et al.} Prog. Theor. Exp. Phys. {\bf 2013}, 03A001 (2013),
and following articles up to 03A011.

\bibitem{evtgen}
 D. J. Lange,
 Nucl. Instrum. Methods, Phys. Res., Sect. A {\bf  462}, 152 (2001).

\bibitem{photos}
 E. Barberio and Z. W\c{a}s, Comput. Phys. Commun. {\bf 79}, 291 (1994);
 P. Golonka and Z. W\c{a}s, Eur. Phys. J. C {\bf 45}, 97 (2006); {\bf 50}, 53 (2007).

\bibitem{geant}
R. Brun {\it et al.} GEANT3.21, CERN Report No. DD/EE/84-1, (1984).

\bibitem{fangks}
%M. Feindt and U. Kerzel, Nuclear Instruments and Methods in Physics Research, Sect. A {\bf 559} (2006)190.
%\bibitem{fangks}
  H. Nakano {\it et al.} (Belle Collaboration), Phys. Rev. D \textbf{97}, 092003 (2018).
  %``$K_S$ selection with NeuroBayes and nisKsFinder class'', Internal Belle Note 1253.

\bibitem{muon}
  A. Abashian {\it et  al.} Nucl. Instrum. Methods. A \textbf{491}, 69 (2002).

\bibitem{electron}
   E. Nakano {\it et  al.} Nucl. Instrum. Methods. A \textbf{494}, 402 (2002).

\bibitem{r2cut}
G. C. Fox, S. Wolfram, Phys. Rev. Lett. \textbf{41}, 1581 (1978).

\bibitem{splot}
M. Pivk and F. R. Le Diberder, Nucl. Instrum. Methods, Phys. Res. Sect. A \textbf{555}, 356 (2005).
%\bibitem{sig}
%R. D. Cousins and V. L. Highland, Nucl. Instrum. Methods Phys. Res., Sect. A {\bf 320}, 331 (1992).
\bibitem{data}
 S. -K. Choi {\it et al.} (Belle Collaboration), Phys. Rev. D \textbf{84}, 052004(R) (2011).
\bibitem{freq}
  G. J. Feldman and R. D. Cousins, Phys. Rev. D \textbf{57}, 3873 (1998).
  \bibitem{sys}
 This expression is only valid if the state is a conventional $c\bar{c}$ state.
%\bibitem{belle2}
%T. Abe {\it et al.} (Belle II Collaboration), 2010, arXiv:1011.0352.
\end{thebibliography}
\end{document}